\documentstyle[12pt,aaspp4]{article}
\begin{document}

\title{Evidence for an accelerating wind as the broad-line region in NGC 3516}

\author{
J.B. Hutchings\altaffilmark{1},
G. A. Kriss\altaffilmark{2,3},
R. F. Green\altaffilmark{4},
M. Brotherton\altaffilmark{4},
M. E. Kaiser\altaffilmark{3},
A. P. Koratkar\altaffilmark{2},
W. Zheng\altaffilmark{3}
}

\altaffiltext{1}{Herzberg Insitute of Astrophysics,
	National Research Council of Canada,\\ Victoria, B.C. V8X 4M6, Canada}
\altaffiltext{2}{Space Telescope Science Institute,
        3700 San Martin Drive, Baltimore, MD 21218; gak@stsci.edu}
\altaffiltext{3}{Center for Astrophysical Sciences, Department of Physics and
        Astronomy, The Johns Hopkins University, Baltimore, MD 21218--2686}
\altaffiltext{4}{Kitt Peak National Observatory,
        National Optical Astronomy Observatories, P.O. Box 26732,
        950 North Cherry Ave., Tucson, AZ, 85726-6732}

\begin{abstract}

    Spectroscopic data in wavelengths 900--3000 \AA\ have been obtained in a
low flux state of the nucleus of the Seyfert 1 galaxy NGC~3516. The line
profiles show P Cygni characteristics, particularly in O VI $\lambda$1032, and
are compared with data from an earlier higher state. The profiles are suggestive
of, and consistent with, an accelerating wind driven by a disk continuum
source, in which the ionisation radii have changed. This scenario may
apply to the formation of other broad emission lines in AGN.

\end{abstract}

\section{Introduction}

While it is now known that roughly 50\% of all Seyfert galaxies
exhibit UV absorption lines (Crenshaw et al. 1999), the Seyfert 1
galaxy NGC~3516 is unique in that its absorption lines are the
strongest and most variable of any object in this class (Koratkar et al. 
1996; Goad et al. 1999). The deep, broad C IV absorption first visible 
in {\it International Ultraviolet Explorer (IUE)} spectra (Ulrich 1988) 
is reminiscent of the broad absorption
lines seen in $\sim 10$\% of radio-quiet QSOs (Weymann et al. 1991).
In many Seyfert 1 galaxies UV-absorbing gas appears in conjunction
with X-ray absorption by highly ionized gas.  These X-ray ``warm absorbers"
are equally common in Seyferts (Reynolds 1997; George et al. 1998).
Crenshaw et al. (1999) note that all instances of X-ray absorption
also exhibit UV absorption.
While Mathur et al. (1994; 1995; 1997) have suggested that the same
gas gives rise to both the X-ray and UV absorption, the spectral complexity
of the UV and X-ray absorbers indicates that a wide range of physical
conditions are present.
In NGC~3516, the UV absorption in the Lyman lines, O VI, N V, Si IV,
and C IV observed using the {\it Hopkins Ultraviolet Telescope (HUT)} requires
multiple zones with differing column densities
and ionization parameters (Kriss et al. 1996a).
Likewise, multiple components with differing physical conditions
are present in the X-ray spectrum (Kriss et al. 1996b).
This inferred complexity is born out by the detection of four distinct
kinematic components in high-resolution spectra of the C IV region using
the {\it Goddard High Resolution Spectrograph (GHRS)} on the
{\it Hubble Space Telescope (HST)} (Crenshaw, Maran, \& Mushotzky 1998).

With observations of only one ion in the GHRS spectrum it is impossible to
determine physical conditions in the absorbing gas.  However, given the
low velocities and widths of two of the C IV components, Crenshaw et al. (1998)
suggest that they arise in gas in the interstellar medium or galactic halo
of NGC~3516.  They associate the more blue-shifted components with outflowing 
gas from the nuclear region. Given these characteristics, we planned
{\it Far Ultraviolet Spectroscopic Explorer (FUSE)} observations of the
900--1200 \AA\ spectrum of NGC~3516 to make high-resolution measurements
of the O VI and Lyman-line absorption components.
As shown by Kriss et al. (2000) for the Seyfert 1 Mrk 509, such
observations can determine physical conditions in individual kinematic
components of the UV-absorbing gas and detect which, if any,
of them might be directly associated with the X-ray absorbing gas.
Ascertaining the physical conditions of the individual components
could give insight into the mechanisms giving rise to the outflow
and hence to the origin of associated UV absorbers in Seyfert nuclei.

For NGC~3516, determining this structure could prove even more enlightening
since it is one of the rare Seyfert 1s whose narrow-line region shows a
bipolar morphology in the distribution of ionized gas around the nucleus
(e.g., Pogge 1989; Aoki et al 1994; Ferruit, Wilson, \& Mulchaey 1998).
In addition, NGC~3516 has an opaque intrinsic Lyman limit (Kriss et al. 1996a),
a feature seen only in the Seyfert 2 NGC~1068 and in only two other Seyfert 1s
with a clear bipolar narrow emission line morphology: NGC~4151 and 
NGC~3227 (Kriss et al. 1997).
In the scheme of unified models for Seyfert galaxies, this is unexpected.
The standard model places the central engine and broad-line region in
the interior of an opaque torus (see the review by Antonucci 1993).
Light from the interior is collimated by the shadowing torus, and it
illuminates the surrounding narrow-line region with a biconical pattern.
Views from the pole of the torus give
a full view of the interior, corresponding to an object we see as a
Seyfert 1. Views in the plane of the torus show an obscured nucleus
dominated by narrow-line emission exterior to the torus, corresponding to
a Seyfert 2.
Thus, both NGC~3516 and NGC~4151 appear spectroscopically as Seyfert 1s,
but their narrow-line morphology resembles that of Seyfert 2s.
This suggests that we are viewing the interior close to the edge of the
collimated pattern, as noted by Hutchings et al. (1998).  
Thus, the outflowing gas we see may be
associated with the more opaque material responsible for collimating the
ionizing radiation, or with an off-axis view of the central engine.

In this paper we describe FUSE observations of NGC~3516 in the
900--1200 \AA\ spectral range and contemporaneous HST observations
at longer UV wavelengths (1150 -- 3100 \AA) using the
{\it Space Telescope Imaging Spectrograph (STIS)} that shed new light
on the outflow phenomenon in Seyfert galaxies.

\section{Observations}

FUSE was designed to obtain high-resolution (R$\sim$20,000) far-UV spectra
covering the 905--1187 \AA\ spectral range
(Moos et al. 2000; Sahnow et al. 2000).
Data are obtained in four independent optical channels, of which two
use LiF coatings on the optics, and two use SiC.
Two micro-channel plate detectors record individual photon events.
A summary of the early FUSE observations of AGN is given by Kriss (2000).
We observed NGC~3516 with FUSE on 2000 April 17, and with STIS
(gratings G230L and G140L) on 2000 April 20.
A full description of STIS and its performance are given by
Kimble et al. (1998).
As the AGN was in an unusually low state, the near-simultaneity of these
observations was of importance (and by design).
We thus have a good set of UV line profiles
in the low state to compare with the normal high state observed earlier
(Kriss et al. 1996a; Crenshaw et al. 1998; Goad et al. 1999).

The FUSE data were obtained in 11 consecutive orbits
through the $30\arcsec \times 30\arcsec$ apertures
for a total exposure time of 16,382 s.
As described by Sahnow et al. (2000), the data were processed using the
standard FUSE pipeline to produce calibrated, one-dimensional spectra
in each of the four channels.
There were two exceptions to this process: first, we restricted the
pulse heights of acceptable data to channels 4--16 to reduce the
background; second, we measured the background in blank regions of each
detector segment and scaled the background constant to this value.
These steps improved the data quality on these faint exposures where
the continuum flux of NGC~3516 is only about
$1 \times 10^{-14}~\rm erg~s^{-1}~cm^{-2}~\AA^{-1}$.
Since the LiF1 optical channel is also the one used by the fine guidance
system on FUSE to maintain a precise pointing, the photometric stability
of the LiF1 data is assured.  In the other optical channels, thermal
drifts can cause mis-centering of the target, leading to loss of flux
and wavelength-scale errors.
We therefore cross-correlated data from the other 3 channels with the
overlapping wavelength range of the LiF1 channel to adjust their
wavelength scales.  We also applied gray-scale corrections to their
flux levels to bring them up to the same value as seen in the LiF1 spectrum.
To achieve a workable signal-to-noise ratio (S/N) while maintaining enough
resolution to recognize and remove H$_2$ absorption lines, the spectra were
binned to a resolution of $\sim$0.5 \AA.
As the different FUSE channels have different S/N levels and residual scattered
light problems, we assembled a complete 905--1187 \AA\ spectrum by selecting
the best data for each wavelength from among the four channels.

The STIS data were obtained in slitless mode.
The G140L spectrum, covering the 1150--1700 \AA\ wavelength range
with a resolution of $\sim1000$, was obtained from a 600 s exposure.
The G230L spectrum, covering the 1650--3100 \AA\ range, was obtained in
a 300 s exposure in the same orbit.
No extended UV emission from the nuclear region of NGC~3516 was visible
in our images.
As the zero point of the wavelength calibration in slitless mode depends on
the location of the target in the field of view, we extracted the
one-dimensional STIS spectra in an iterative process.
Using the STIS calibration pipeline in the Space Telescope Data Analysis
System (STSDAS) {\tt stis} package, we first obtained a preliminary spectrum.
We measured the wavelengths of foreground Galactic absorption features, and
used these to determine appropriate offsets for the centering of the target.
We edited these offsets into the headers of the two-dimensional flat-fielded
data frames (as is usually done by the wavelength calibration process in the
pipeline) and re-ran the data through the pipeline to obtain our final
calibrated spectra with corrected wavelength scales.
Compared to the FUSE spectra, the STIS spectra are extracted from a much
smaller angular region surrounding the nucleus.
They have somewhat worse resolution but higher S/N.
In the overlap region from 1150 to 1180 \AA, the FUSE and STIS continuum levels
match quite well, so it appears that the continuum arises almost entirely in
the nuclear region with little to no contribution from starlight at larger
radii in the FUSE aperture.

The diagrams illustrate the main features in the new spectra.
We also compare our new spectra with data of lower resolution from HUT,
which observed NGC~3516 in a high state in 1996 (Kriss et al. 1996a).
For comparison of the whole FUSE
wavelength range, the FUSE data were further smoothed to match the HUT data.
As the overall FUSE signal level was low (due to the target faintness and
observing time available),  it was not
possible to obtain detailed spectra of the Lyman lines and Lyman limit in
NGC~3516.  Therefore, the analysis and discussion below are not
as detailed as would be warranted by better S/N data.

\section{Line profile changes with nuclear flux changes}

   Figure 1 shows plots of lines from our data compared with earlier,
high state, data. For easier comparison, the plots shown were placed on a
velocity scale, centred at the galaxy redshift for each line. This is
intended only to show the extent of the line asymmetries, which is not 
very sensitive to the range of uncertainty in the true redshift value.
The 1996 data 
were also scaled to match the 2000 low-state continuum, for better 
comparison of the line profiles. This scaling is simply a factor applied
to the entire flux so that the continuum levels beyond the emission
lines match to within about 10\%.

The O VI velocities are given with respect to redshift 0.0091. This is 
higher than the published value of 0.008836 and is derived from the 
reflection in velocity space that achieves the most symmetrical 
outer broad emission profile for O VI. The other lines shown used the 
published redshift value but were sufficently asymmetrical for various 
reasons that they were not useful for defining the redshift. There is
uncertainty in this approach for O VI and the velocities in its plot are thus 
uncertain by some 50 $\rm km~s^{-1}$. The ubiquity of the shortward
absorption in O VI, Ly$\alpha$, and C IV in this current low state of 
the galaxy, and the strong narrow absorptions in all states, makes it
hard to define the line centre precisely.

   The top left FUSE O VI profile in Figure 1 is the combination of both 
O VI lines, after approximate removal of the H$_2$ absorptions, which were
recognised from FUSE
data of different targets with high S/N spectra. The plot at lower right
shows both profiles without H$_2$ removal. The lower right plot also 
shows the HUT data on the same flux scale. The doublet ratio of the O VI
line peaks in 2000 is nearer to unity than in the HUT 1996 data, indicating 
that the optical depth was greater in 2000.

   The dashed lines on the left of the profiles are the reflected longward
profile assuming the above velocity zero. They thus show the extent of the
implied absorption in the profiles. The HUT Ly$\alpha$ profile is
heavily contaminated by local absorption (and geocoronal emission) to its
shortward side beyond -1000 $\rm km~s^{-1}$ or so.

    The FUSE and STIS data are essentially simultaneous (obtained on
April 17 and 20, respectively). For comparison,
the dotted profiles are the HUT data from 1996, when the nucleus was in a
higher state. For easier profile comparison, the HUT data have been roughly 
scaled to match the 2000 continuum levels. Thus, the flux scale shown applies
only to the year 2000 FUSE and STIS data. The changes in the relative 
absorption between 1996 and 2000 are clear from these comparisons.
Note that the C IV absorptions seen earlier are still present, but that
there is extra absorption at shorter wavelengths in the 2000 low state.

   The upper right profiles show the broader Mg II and C III] profiles
(the latter blended on its shortward side by Si III] 1892\AA). 
As noted by Goad et al. (1999),
these lines show no changes or absorption from high to low state. The 
comparison between C IV and Mg II in centre right shows that the C IV
broad emission profile matches the Mg II, with the
absorptions lying below this broader profile.

    Figures 2 and 3 show the full wavelength ranges of FUSE and the FUV
channel of STIS, where other lines can be compared between the high and
low states of NGC~3516. Similar changes are seen in Si IV, while the weaker
line changes are more noisy. The large change in He II $\lambda$1640 is 
probably due
to the larger aperture in the HUT data including hot stars as well as the
active nucleus. The Lyman absorptions seen in the HUT spectrum are not seen
in the low-state FUSE spectrum. However, there is severe airglow 
contamination in Ly$\beta$, and the signal is noisy and weak at the higher 
Lyman lines. 

\section{Low state line profiles}

  The broad line profiles (both emission and absorption) 
in the new low-state data strongly suggest that they
are dominated by outflow at velocities not seen in the high state. 
The absorption and line width in O VI indicate
outflow velocities of some 600 $\rm km~s^{-1}$ in the hottest (highest
ionisation) regions, with the bulk at about 400 $\rm km~s^{-1}$.
There is absorption (outflow gas) at all velocities down to zero.
The C IV and L$\alpha$ profiles show the same velocity of maximum absorption
and absorption to zero velocity, but the outflow occurs to higher velocity,
with a second minimum at 1600 $\rm km~s^{-1}$ and a terminal velocity of
2000 $\rm km~s^{-1}$. Much, but not all, of the low-velocity
outflow is also present in the high-state profiles, and we know that these
can be resolved into several sharp components, from higher resolution data
from HST. 

The lower resolution of the STIS spectra means that we cannot easily see
whether there are sharp components to any of the current outflows. The HUT
data on O VI have very poor resolution and noise but suggest a different O VI
optical depth overall. We also see that the emission from the low ionisation 
and low density gas (C III] and Mg II) are the same as in the high state. 
Thus, we seem to be seeing a change in the state of the high-ionisation gas.
As it coincides with continuum flux decrease, we speculate that there is
increased absorption in the inner part of the BLR responsible for the overall
phenomenon.

   The scenario that we suggest for the broad emission and broad absorption
features is of an accelerating wind like a stellar
wind. The velocity increases and the ionisation decreases outwards. As the
central source flux drops, the ionisation boundaries move inwards. Wide
profiles with weak or non-detectable P-Cygni absorptions arise if lines are
formed in a large radius region at terminal velocity. This is the situation
in the high state (assuming the narrow low velocity absorptions arise in places
that lie far outside the wind region). In the low state, the high ionisation
lines  are formed at much smaller radii from the nucleus, where P-Cygni
absorptions are more prominent, and can be seen through the whole absorbing
column to the continuum source. The highest ionisation lines are narrower
and have lower velocity absorptions. We describe and quantify this
scenario in a little more detail in the next section.

   We note that the C IV profile also contains strong low
velocity absorption, in both high and low states. This is resolved in
higher resolution HST data into several narrow features that do not
change, and are similar to narrow absorptions seen in other Seyferts
(Crenshaw et al. 1999). These arise in the general ISM and stable outflows 
that lie well outside the BLR (as seen also in NGC~4151 by Kaiser et al. 2000),
and thus need to be
ignored or our discussion of  the BLR profiles. The FUSE resolution is high
enough to show that the P Cygni absorption seen in O VI is not made of narrow 
features, but is broad, as expected from the wind model.

   We note that the two higher velocity C IV absorptions resolved in high
resolution HST spectra have velocities that are close to those seen in the
FUSE O VI profiles. While the O VI profiles are not resolved into the
distinct components seen in C IV, it is possible that the absorbing
material is the same if the O VI absorbers are much more saturated.
If this is so, then we require a different explanation in terms of variable 
high ionisation of dense material outside the BLR. It does not explain the
narrowness of the O VI emission in the FUSE data. Thus, while acknowledging 
the possibility that the lower velocity absorbers have a different origin,
we will discuss in more detail the disk wind idea that appears to
explain more of the observed behaviour.

\section{The wind scenario for the BLR}

   The profiles are broadly consistent with an accelerating wind, as present
in OB stars. It is worth checking this idea for consistency, since it
implies a very specific model for the BLR. Figure 4 outlines the scenario
in cartoon
fashion. The ionised gas that constitutes the BLR is accelerated away
from the central continuum source until it approaches a terminal velocity. The
terminal velocity is indicated by the overall width of the emission lines, 
and where there are P-Cygni absorptions, by the maximum blue-shifted absorption.
There is an ionisation (temperature) gradient outwards from the central
source. Line profiles are formed in the range of radius (hence velocity)
where its ion is found. Thus, lines are formed in shells of increasing
radius (and velocity) as the ionisation state decreases. 

   The cartoon profiles show qualitatively how the profiles differ as 
the shell of formation
moves outward. These hold if the shells lie within a few times the radius of
the continuum source. Thus, the absorption column is a decreasing fraction of
the projected emission line region, as the shell radius increases. Close to
the continuum source, there is strong absorption, and all velocities are 
low, so that the overall profile is narrow. It is also relatively `peaky'.
For shells of increasing radius, the overall profile width increases, and the
absorption becomes weaker and occurs only at the shortward edge of the
profile. Once terminal velocity is reached, the profile width remains the
same, and the absorption becomes negligible at larger radii. The details
of the profiles depend on the exact relationship between ionisation, radius,
and velocity. 

    We note that this simple model is for spherically symmetric expansion, as
expected for a star. If the wind is driven by a disk, the geometry becomes
a further parameter, and the absorption and emission components may be
separated if the rotation velocity is comparable to the outflow velocity.
Knigge and Drew (1996) discuss 
disk winds seen in cataclysmic binary systems, that show some of these
effects. Disk winds may be driven over a broad angle on each side of the 
disk, but not near the disk plane. Since AGN jets are also driven perpendicular
to the disk, there may be a cone of wind avoidance where the jet lies, if 
there is one. The Seyfert 1 paradigm is that we view the central disk and BLR
directly without obscuration by the opaque torus, and thus within the 
expected sightline of a central wind. The wind profiles for
such a disk sightline are similar to those for a spherical wind, with slightly 
less peaky emission profiles. Thus, the simple wind cartoon is generally
applicable to the expected geometry, but dependent in detail on the
solid angle filled by the wind. 

We note that the disk wind model of
Murray et al. (1995) for BALQSOs blows close to the disk
plane for objects of higher luminosity and outflow velocity (see also
Elvis 2000).
The hydromagnetic wind of K\"onigl and Kartje (1994), on the other hand,
is more collimated normal to the disk plane. Proga, Stone, and Kallman (2000)
discuss the dynamics for disk winds from more massive central objects,
and note the role of X-ray shielding and ionisation that launch and
accelerate the wind in such models.   
The strong X-ray flux from the nuclear source is likely to highly ionize
unshielded portions of the disk wind, as emphasized by Murray et al. (1995).
This highly ionized portion of the wind could be associated with the X-ray
warm absorber.  Mathur et al. (1997) discuss X-ray warm absorbers in
NGC~3516 in the context of an outflowing wind, but, unfortunately, we do not
have contemporaneous X-ray spectra that would permit us to test their
predictions.

   We now postulate that the change between high and low states of NGC~3516
are principally expansion or shrinking of the ionisation radii within the
wind, as shown in the sketch. This would arise from changes in the amount of
ionising radiation that are effectively the observed high and low luminosity
states of the nucleus. The changes seen in the profiles of high, intermediate,
and low ionisation, as illustrated by O VI, C IV, and Mg II, are much as
observed. 
The postulated stratification is both a function of ionisation state as well
as the line formation mechanisms in the wind.  Both thermal emission and
scattering contribute to the formation of wind emission lines.  As in OB
stars, the low-ionization lines are likely dominated by thermal emission,
while scattering will dominate the observed high-ionization line emission.
Since scattered line profiles are sensitive to the velocity gradient,
broader lines are formed are larger radii.
Thus, the overall idea seems feasible, \it provided that the BLR 
lies within a few disk radii of the continuum source. \rm

   We can also imagine that the high to low state changes involve changes 
in the continuum source radius or the velocity profile within the BLR. 
The flux changes are no more than a factor of 2--3, so that we do not expect
disk-size or wind-acceleration changes to be large or rapid. The change of 
the ionisation radii will be the immediate and dominant effect in any case.

   We can make a rough estimate of the radii, based on the assumption of
similarity with OB stellar winds. The stellar winds are driven by the radiation
pressure in the rest-frame UV where the strong absorptions lie that drive 
the wind
by photon scattering. The nucleus of NGC~3516 has similar colour to an OB star,
so we may compare the magnitudes to estimate the radiation that drives the
winds. NGC~3516 has a redshift of $\sim$2650 $\rm km~s^{-1}$ which gives it a
distance of some 40 Mpc or distance modulus of about 33.
The galaxy V magnitude is 12.4, (with the nucleus at 14 - 16)
which yields an absolute magnitude of -20.6, of which the nucleus accounts
for about -18. An OB star with a strong wind has absolute V magnitude of -7
(with similar colours), so that the nuclear continuum source is equivalent to
some 60,000 OB stars. Assuming we require the same radiation level for the
disk wind of NGC~3516, this implies a continuum source radius of 250 times
that of an OB star, which is about 60 a.u. The BLR then should have a radius
of several times this, which is about 2 light days in radius. Echo mapping
observations indicate BLR diameters of a few light days for Seyferts. 
In the case of NGC 3516 it is measured at 4.5 days for the total C IV emission
line flux (Koratkar et al 1996). Thus, within the uncertainties of the
luminosity estimate above, the
scenario seems very consistent with a radiation-driven wind for the BLR. In
fact, the numbers above would \it require \rm a wind to be present by the same
processes that we understand for stellar winds.

  It is interesting to see how the masses and mass flows scale in this
rough comparison. Stellar winds have terminal velocities closely related to
the surface escape velocity of the star. The terminal velocity in NGC~3516
is $\sim$2000 $\rm km~s^{-1}$,
very typical of stellar winds, so that if the mechanisms
are the same, the average effective gravity at the base of the AGN wind 
should be comparable with an OB star. The average particle at the surface of 
the AGN accretion disk (of radius 250 OB star radii) lies at distance
$\sim$200 OB star radii from the central object. To make the gravity
equal in both cases requires a central mass of about 10$^6$ M$_{\odot}$,
which again is the right order of magnitude. If we scale the mass loss
against typical stellar winds of about 10$^{-6}$ M$_{\odot}$ per year, 
allowing that
the AGN wind may fill say 1/4 of a sphere, we find a mass-loss rate of
a few M$_{\odot}$ per year. This again is not unreasonable, but would be
significant in the accretion budget for the AGN.

In in more comprehensive picture, we note that ionisation by high energy
photons from the AGN will alter the wind as its opacity changes with height,
probably giving rise to the warm absorbers seen in the X-ray spectra. The
wind structure that may apply in NGC~3516 will need detailed modelling, and
comparison with data from different luminosity states will help define the
important ionisation structure. Such modelling---particularly the dynamics
of rotation and acceleration of the wind--- must await more and better
contemporaneous data at FUV and X-ray wavelengths.

   Our principal conclusion from the present data is that we feel there is
significant evidence that the BLR in NGC~3516 is an accelerating wind driven 
by the nuclear continuum source, that is seen in sightlines within the
cone of UV ionisation. If this applies in a more general way, we might
expect to see similar profile changes in other Seyferts with very low 
states. In the case of NGC~4151, this does seem to be true: P Cygni profiles
are seen in lines in low states - most notably in data taken in the current
(year 2000) low state by STIS. We therefore suggest that careful monitoring
of line profiles with nuclear flux would be a valuable way of verifying the
model proposed and learning more about how it works. 

Our discussion relates to the connection between broad-line profile
widths and the mass of the central black hole. Clearly, any motions in the
central region are dominated by the gravitational field, whether they be
in Keplerian orbit, free-falling, moving with escape velocity, or driven
by radiative forces. More detailed wind models may enable us to use
profile widths to test them, and thereby formulate the relation between
observed profiles and central mass. We note that the low state of NGC 3516
we have sampled, may reveal the inner part of the outflow not often seen in
more luminous and extended BLRs, where line widths are no longer radiatively
driven.  

 Finally, we note that the highest ionisation lines, such as the Fe K line 
seen at 7Kev in X-ray spectra, have profiles that show they arise in a 
disk of relativistic velocities, so do not arise in the BLR wind, but in 
the continuum source disk itself (Nandra et al. 1999).

\section{Other spectral features}

   Figure 2 shows the entire FUSE range spectrum compared with the same
range with HUT in 1996. For this plot, the FUSE data were selected to use only
the best data for each wavelength region, and then smoothed to about 0.4 \AA\
resolution, for easier comparison of the broad weak features. The airglow lines
have been removed, but not the H$_2$ absorptions, which must be present
in both datasets. The HUT data also show zero redshift L$\beta$ absorption
which has not been removed. At the shortest wavelengths, there are probably
detector and noise effects that are not real differences.

The lines present other than O VI are N III $\lambda$990, He II
$\lambda$1085, and possibly N IV $\lambda$948
and S IV+Si IV $\lambda$1063-66. The C III $\lambda$977 line is not detectable,
while the other lines are all narrower as well as weaker than in the 1996
spectrum. There are very broad features that appear to be associated with the
lines of C III, N III, O VI, and He II that are seen at both epochs, for which
we offer no detailed explanation. The overall continuum level in 2000 is 
about 1/3 that of 1996, and the broad features all scale with this same factor.
There are differences at the positions of N III and the S IV/Si IV blend.
Sharper emissions are seen on top
of the broad wings at He II (only in 2000) and N III and O VI.

   Moving to longer wavelengths (1200 to 1800 \AA), Figure 3 shows the 
comparison between the 1996 HUT data and the 2000 STIS data. The flux level
difference becomes smaller with increasing wavelength and is close to
2 at 1800 \AA. Aside from emission line profile changes, the relative
strength of He II and N III] are lower in the 2000 low state. Since it is
unlikely that the forbidden line flux changes significantly, the difference is
probably due to the line emission arising over an extended region that is
differently sampled by the STIS and HUT slits. The FUSE and STIS continuum
levels match quite well where they join, so that it appears that the continuum
arises almost entirely in the nuclear region. The drop in He II $\lambda$1640
flux between HUT and STIS may also indicate an extended origin of this emission,
since the He II $\lambda$1085 line does not show the same drop in flux.
He II $\lambda$1640, along with He II $\lambda$4686, arises in hot stars and
thus may well be extended in this galaxy.
Thus, among the weaker emission lines from the nuclear region, both N III
and He II have developed sharp components in the 2000 low state.

\section{Discussion}

      The paradigm of the unified model for AGN is that in Seyfert 1
galaxies, we view the
broad line region from within the cone of light illuminated by the nucleus.
The cone is caused by shadowing of an equatorial region by an opaque torus
around the nucleus: Seyfert 2s are those nuclei hidden by the torus.
Inside the cone there is ionising radiation, radio jet material, and clouds
of narrow emission line material. Recent data from STIS (e.g. Hutchings
et al. 1998, Kaiser et al. 2000, Crenshaw et al. 2000) suggest that the
narrow-line material is moving outwards at a few hundred $\rm km~s^{-1}$ along
the inside surfaces of the cones, and not filling them.
The STIS data indicate there are also faster moving clouds that fill a wider
angle and arise close to the nucleus (Hutchings et al. 1999). These
are seen both in absorption and emission, and the former are usually stable
over years. Occasionally some narrow absorptions change or new ones appear.

   In addition to these narrow and relatively stable absorbers,
the change of broad line profiles from 1996 to 2000 suggest that they
may be explained by an outflowing BLR. This paper has explored the
way in which the change of state allows us to probe the
velocity structure of a broad-line wind, by varying the radii of the shells 
in which different lines are formed. We think that the density structure may 
not change significantly. The geometry of the wind, what happens to its
material as it moves outward and cools, and its relationship with the other
mass flows, need to be considered and tested with observations in other objects
as well as NGC~3516 itself as it continues to vary. Our scenario predicts
a close link between the BLR profiles and the continuum flux, which can be done
with continued monitoring, particularly in the O VI lines.

   Kriss et al. (1996a) note that the presence of Lyman limit absorption
and the S-shaped NLR in NGC~3516 may imply a line of sight near to the
edge of the cone, as in NGC~4151. This does not preclude the disk wind
scenario, but it would be useful to study the NLR dynamics in detail
for more geometrical clues, and obtain better FUSE data to look for the
Lyman absorption lines and limit.

   There have been other discussions of the BLR of NGC 3516, based
on monitoring campaigns for echo-mapping, as well as more sporadic
observations of high and low states and the occasional presence of
BAL-type absorption (e.g. Wanders et al 1993, Koratkar et al 1996, Goad et al 
1999). The suggested model in this paper is consistent with these other
discussions, none of which have proposed a detailed model. 

     NGC 3516 has a NLR component, which we assume does not change with
the nuclear and BLR changes, since it is formed in the much more extended
S-shaped region noted above. This component is not seen in the UV lines
in Figure 1, although Goad et al (1999) model profiles with a narrow component
of FWHM some 1200 km s$^{-1}$. The narrow emission peak seen in O VI is much
narrower than this, and is unlikely to represent the NLR emission for this
reason as well as for its high ionisation. Aoki et al (1994) also discuss
the ionisation of the NLR and its relationship to the nuclear source.

   Our FUSE profiles of O VI are key to our suggested model, and we have
noted that a higher S/N would be very useful in checking whether these lines
do contain NLR components such as the narrow absorbsions and emission peak.
With the data in hand we favour the explanation that they arise in the
inner accelerating part of a central mass flow.  

   There are some clues in other objects. VLBI radio maps of NGC~4151 
shows a bend close to the nucleus (Roy et al. 1999), where the radio
structure becomes aligned with the outer ionisation cones and radio structure.
The same galaxy also shows P Cygni profiles occasionally, and notably in its
current low state.
In NGC~4151, our line of sight is deduced to be close to the edge of the
cone.

   The relation between line FWHM and reverberation time lag (Peterson and
Wandel 2000) for other AGN, indicates that in those cases the line profiles
become narrower with distances from the nucleus that are larger than those
we consider here. This would indicate that the wind velocity falls again at
radii where the radiation pressure has dropped and the wind material is slowed
by interactions with the ambient medium. It is only in cases such as the low
state of NGC~3516, that broad line formation is dominantly in the inner
accelerating region.  

   We may further speculate whether a wind model for BLR applies to more
energetic QSOs. These have lines (and BAL velocities) that are higher
by a factor of several, luminosities higher by up to 100 or more, BLR sizes
(from variability timescales) larger by about 100, and central masses 
probably 2 to 3 orders of magnitude higher. These numbers all scale quite 
reasonably to winds, although the above parameters are not tightly linked
to each other. Good resolution and S/N studies of variations in highly 
ionised emission lines in a range of AGN will be necessary to pursue these
questions.

This work is based on data obtained for the Guaranteed Time
Team by the NASA-CNES-CSA FUSE mission operated by the Johns Hopkins University.
Financial support to U. S. participants has been provided by NASA contract
NAS5-32985.

\clearpage

\centerline{Figure captions}

1. Comparison in velocity space with respect to the galaxy velocity, of
profiles from 2000 low state with 1996 high state. In the left panels,
the dashed lines show the reflected longward profile to highlight the 
shortward absorption. The top left O VI profile combines both lines of 
the doublet, with the H$_2$ absorptions removed. The full O VI data are 
shown in lower right panel, with H$_2$ lines
marked with vertical dashes. The 1996 fluxes 
have been scaled to match the continuum levels from the low state, to show 
the profile differences. FUSE data are smoothed to 0.5 \AA\ resolution and the
HUT data have 2 \AA\ resolution. Note the increased broad absorption in
the low state data.   

2. Comparison of the low and high state spectra over the entire FUSE range.
The FUSE data are smoothed to match the HUT data, and the HUT data are 
displaced in Y by 2x10$^{-14}$ (dashed line) for clarity.
The principal broad emission lines are identified above the HUT spectrum.
Other broad features with possible identifications are marked where 
they are seen 
in the FUSE data. In addition to the strong O VI, N III and He II
emission peaks, there appear to be broad components that scale with
the continuum. Overlapping STIS coverage is shown as dotted line.

3. Comparison of the STIS FUV spectral range with the HUT high state spectra.
Note the profile changes in Si IV, and He II as well as those shown in
Figure 1.

4. Sketch showing the way that a contraction of the ionisation structure in the
low state can explain the profile changes seen. While no units are given,
radii range from the surface of the continuum source to several times its
radius. The text discusses this diagram in detail.

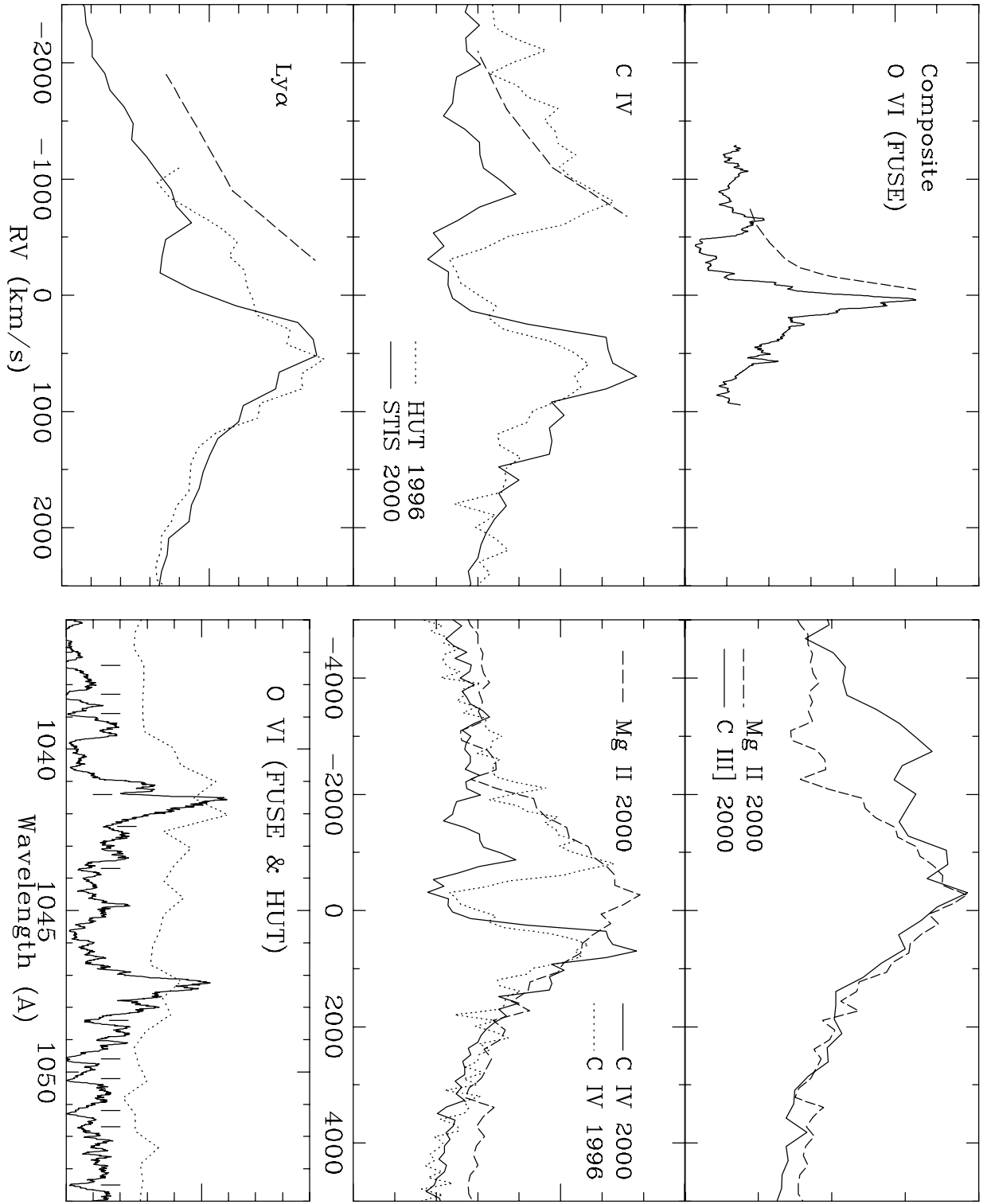
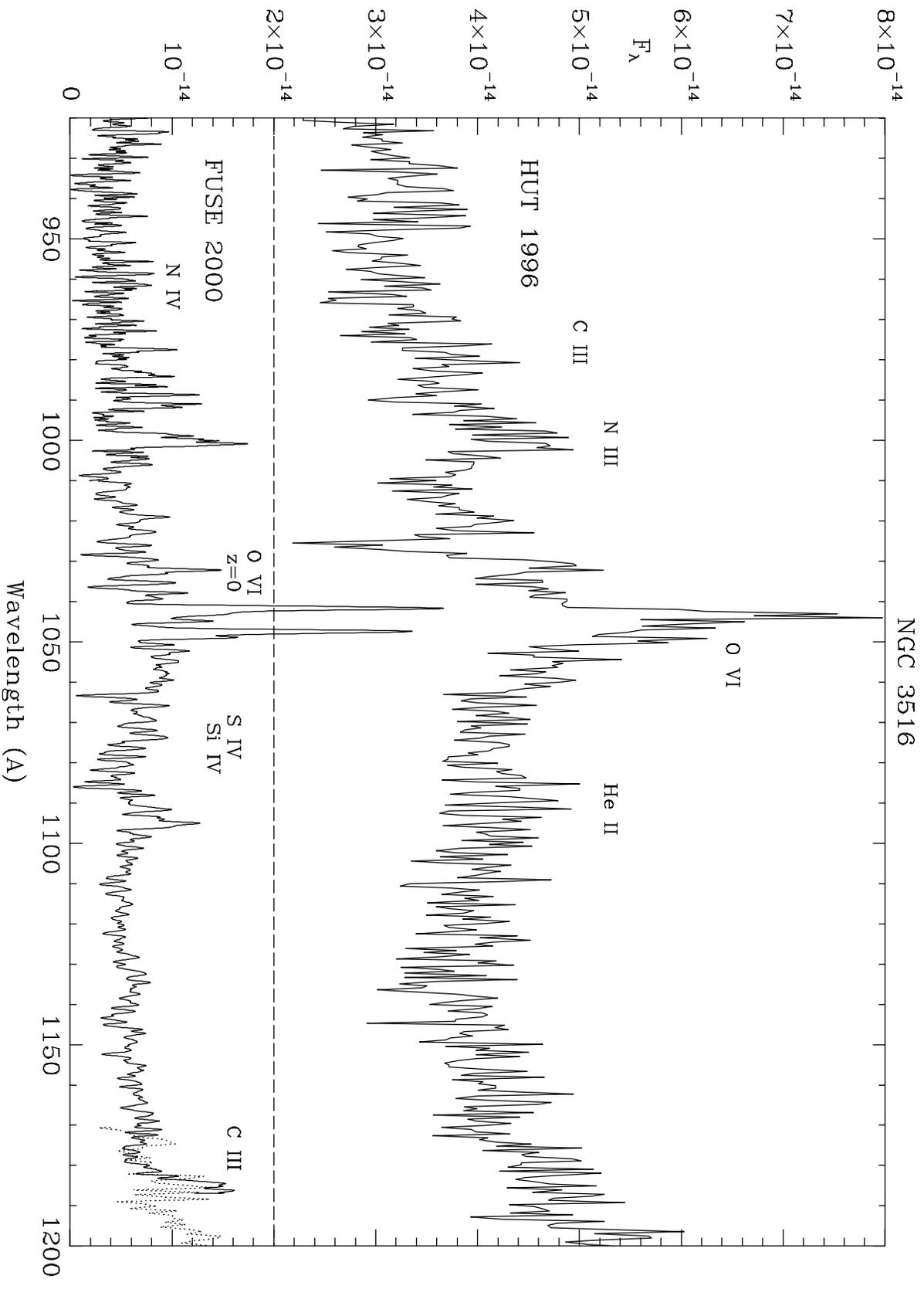
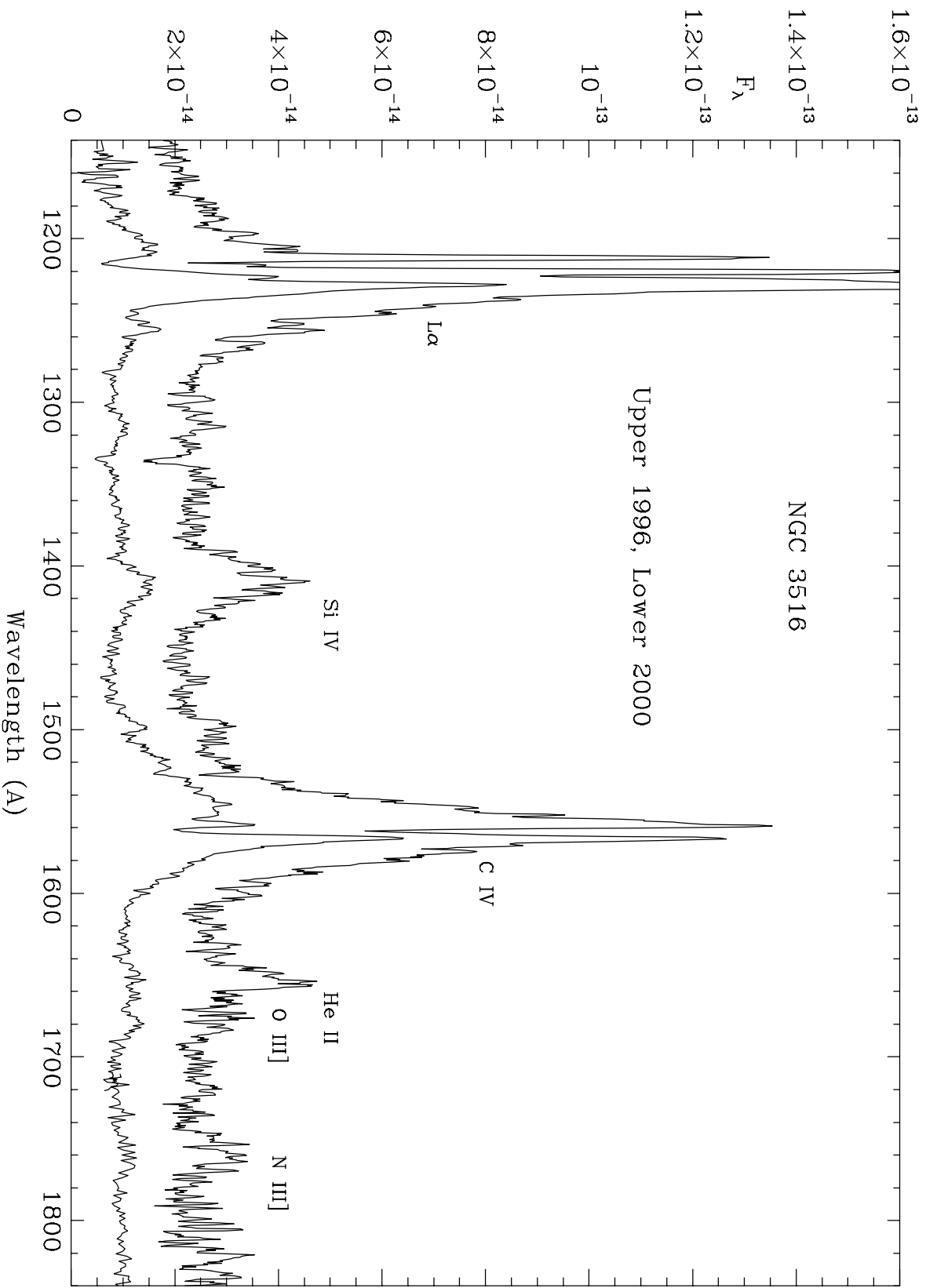
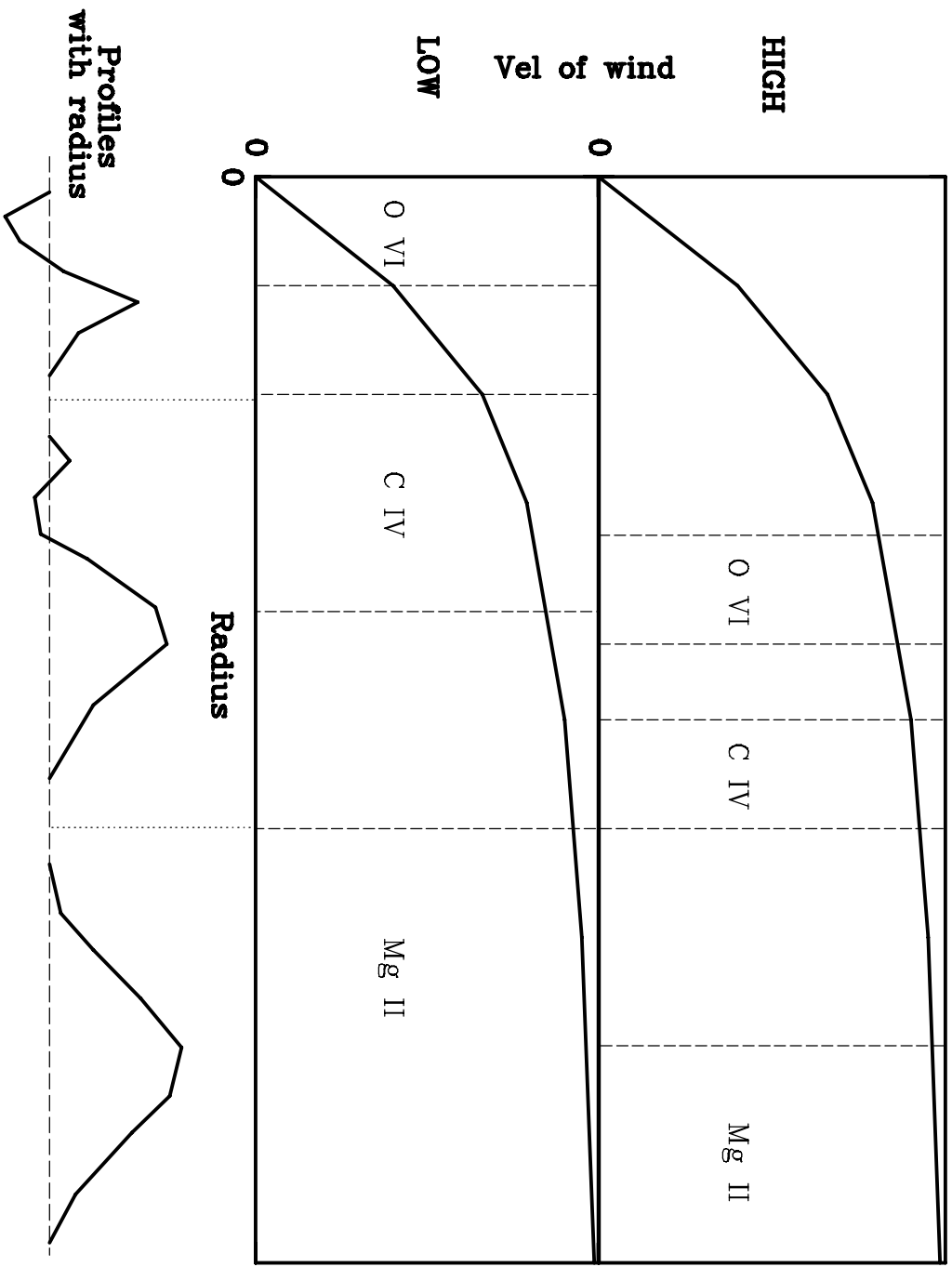
\end{document}